\documentclass[prl, twocolumn,superscriptaddress,preprintnumbers,a4paper,amsmath,amssymb,showpacs,floatfix]{revtex4}
\usepackage{graphicx}
\usepackage{color}\color[rgb]{0.000,0.000,0.000} 
\usepackage{amssymb} 
\usepackage{amsmath} 
\usepackage{multirow}

\begin{document}
\newcommand{\mn}{MnS$_{2}$}
\newcommand{\te}{$_{t}$}
\newcommand{\ot}{$_{o}$}
\newcommand{\Tc}{T$_{C}$}
\newcommand{\Ts}{T$_{s}$}
\newcommand{\Tn}{$T_\mathrm{N}$}
\newcommand{\MuB}{$\mu_\mathrm{B}$}
\title{Unifying local and average structure in the phase change material GeTe}

\author{Jessica M. Hudspeth}
\altaffiliation{Presently at: Universit\'e Pierre et Marie Curie, Paris, France.}
\affiliation{European Synchrotron Radiation Facility (ESRF), 71 avenue des Martyrs, BP 220, 38043  Grenoble Cedex 9, France}

\author{Tapan Chatterji}
\affiliation{Institute Laue Langevin (ILL), 71 avenue des Martyrs, BP 220, 38043  Grenoble Cedex 9, France.}

\author{Simon J.L. Billinge}
\affiliation{Department of Applied Physics and Applied Mathematics, Columbia University, New York 10027, USA.}

\author{Simon A. J. Kimber}
\email[Email of corresponding author:]{kimber@esrf.fr}
\affiliation{European Synchrotron Radiation Facility (ESRF), 6 rue Jules Horowitz, BP 220, 38043  Grenoble Cedex 9, France}

\date{\today}

\begin{abstract}
The prototypical phase change material GeTe shows an enigmatic phase transition at $T_{C}\sim$650 K from rhombohedral ($R3m$) to cubic ($Fm\bar{3}m$) symmetry.  While local probes see little change in bonding,  in contrast, average structure probes imply a displacive transition. Here we use high energy X-ray scattering to develop a model consistent with both the local and average structure pictures. We detect a correlation length for domains of the $R3m$~structure which shows power law decay upon heating. Unlike a classical soft mode, it saturates at $\sim$20 \AA~above $T_{C}$. These nanoclusters are too small to be observed by standard diffraction techniques, yet contain the same local motif as the room temperature structure, explaining previous discrepancies. Finally, a careful analysis of the pair distribution functions implies that the  0.6 \%~ negative thermal expansion (NTE) at the~$R3m$~-$Fm\bar{3}m$ transition is associated with the loss of coherence between these domains.
\end{abstract}
\maketitle
\noindent{\textit{Introduction:}} Materials which can be rapidly switched between states with contrasting properties are useful for memory applications \cite{memory}. Examples include the chalcogenide phase change materials which are now making their way into consumer devices. These materials can be switched between crystalline and amorphous states using laser heating on a nanosecond timescale \cite{switch}. The large change in conductivity, combined with kinetic trapping of the amorphous state, allows them to be used as two-state bits. Understanding the mechanisms of phase transitions in these materials is thus of key importance. However, a cursory review makes it clear that their simple structures and compositions belie structural complexity. When doped with antimony, GeTe produces one of the most promising phase change materials, yet this simple parent phase is still not fully understood despite decades of research. In common with many other pnictides and chalcogenides, strong chemical constraints on bond lengths are found. It is this interplay between covalent bonding, coherent lattice vibrations and electronic properties which makes these materials so interesting. In particular, GeTe shows a second order Jahn-Teller  (SOJT) distortion~\cite{Peierls}~of the first coordination sphere, due to the formation of Tellurium lone pairs. At room temperature, this is manifested as an ordered arrangement (Fig. 1a) of three long and three short Ge-Te bonds in each GeTe$_{6}$~octahedra, this causes a rhombohedral distortion. Upon heating however, the rhombohedral angle decreases in an order parameter-like manner~\cite{Tapan1}, and a structural phase transition is found near 650 K. The resulting cubic rock salt structure imposes a single symmetry equivalent Ge-Te distance. Average structure probes thus imply that the transition is of second order.  Spectroscopic measurements lend weight to this interpretation, for example inelastic neutron scattering \cite{Tapan2}~and phonon calculations show that a triply degenerate phonon condenses at $T_{C}$. Similar observations have been made with Raman scattering \cite{Raman}. In great contrast however, pioneering local structure measurements \cite{EXAFS} using EXAFS showed that the  Ge-Te bond lengths are essentially unchanged upon heating up close to the melting point. In fact,  the local SOJT distortion even persists into the liquid phase \cite{melt}. Unifying the local and averages structure in GeTe is thus an important and thus far unsolved problem. Here we use high-energy x-ray scattering to address this question through so-called total scattering. This technique utilises both the coherent Bragg diffraction as well as diffuse scattered x-rays, to calculate a pair distribution function (PDF) and is thus sensitive to both local and long range order. This technique is also more sensitive than EXAFS to $intermediate$~range order on the nanometer length scale. One earlier report applying the PDF technique to GeTe exists \cite{PDF}. This short work confirmed the presence of two Ge-Te distances in the high temperature cubic phase. In this Rapid Communication, we re-visit GeTe as a function of temperature, and apply full structure refinements over a range of length scales. We show that: i) the 650 K transition corresponds to the loss of coherence between nanoscopic distorted domains. These are too small ($\sim$ 20 \AA) to be observed by diffraction methods, and persist close to the melting point. ii) We show that the negative thermal expansion \cite{Tapan3}~at the transition is likely driven by increased dynamical disorder, as the average atomic distances distances within the distorted domains are essentially unchanged.
\begin{figure}[tb!]
\begin{center}
\includegraphics[scale=0.39]{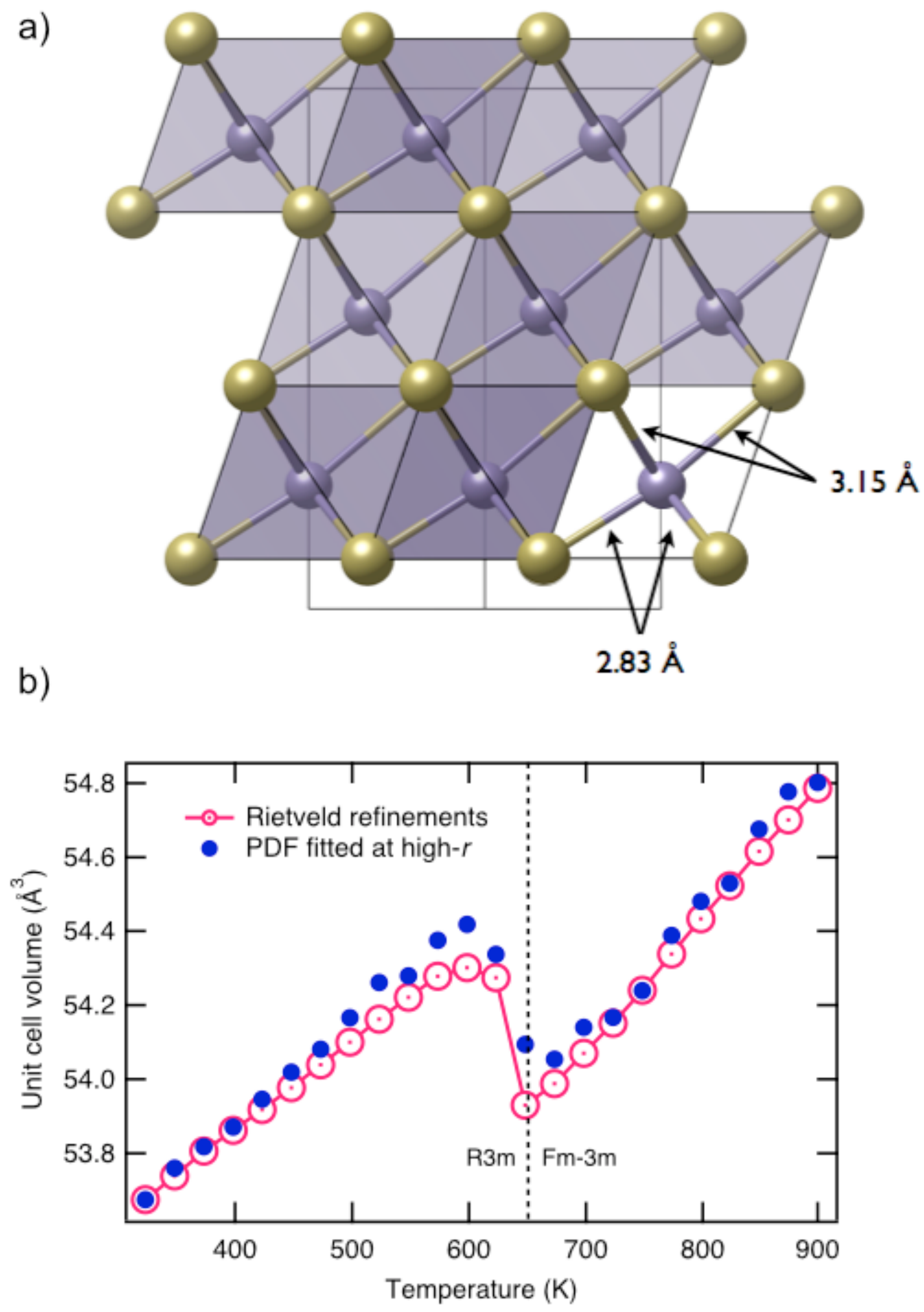}
\caption{(color online) a) Structure of $R3m$~GeTe. The purple spheres represent the Ge atoms, while the golden spheres represent Te atoms. One GeTe$_{6}$~ octahedra is shown in cutaway, emphasising the off-centre distortion which generates a long and a short Ge-Te distance; b) Refined reduced unit cell volume for GeTe as a function of temperature. The open symbols are results from our Rietveld refinements, while the closed symbols are from real-space refinements against experimental pair distribution functions in the range 30$<r<50$ \AA~as described in the text.}
\label{Fig1}
\end{center}
\end{figure}
\noindent$Experiment:$ Our sample of GeTe was the same as used in the original neutron diffraction investigation \cite{Tapan1}. A small piece of the crystal boule was ground into a fine powder for the measurements reported here. The sample was sealed in a quartz capillary and diffraction experiments were performed on the ID15B high energy x-ray scattering beam line of the European Synchrotron Radiation Facility (ESRF), in Grenoble, France. The incident energy was 87 keV, and the scattered x-rays were detected with a Mar345 image plate. This was placed close to the sample to obtain data reaching high momentum transfers ($\sim$30 \AA$^{-1}$). At each temperature point, multiple frames were collected and averaged in the close configuration, before the detector was translated further away to collect a data set suitable for conventional Rietveld analysis as in previous investigations~\cite{Li2RuO3}. Two dimensional data were azimuthally integrated~\cite{pyFAI} using pyFAI, and the PDFGetx3 and PDFgui packages were used to extract~\cite{PDFgetx3} and model~\cite{PDFGui} the pair distribution functions respectively. The structure factors were truncated at $Q_{max}$=26 \AA$^{-1}$~before Fourier transformation. Rietveld analysis was performed \cite{GSAS,EXPGUI}~using GSAS and the EXPGUI interface.\\
\noindent$Results:$ Turning first to the results of the Rietveld analysis, in agreement with earlier reports, a clear rhombohedral splitting was observed at room temperature. Upon heating to $\sim$ 650 K, the splitting reduced until a metrically cubic lattice was obtained \cite{Tapan1,Tapan3}. Also in agreement with earlier results, the reduced unit cell volume shows a $\sim$0.6 \%~collapse at the R-C transition (Fig. 1b). As a further check on the high temperature symmetry, we also collected a data set in the cubic phase using the crystal analyser diffractometer ID31, also at the ESRF. This instrument~\cite{ID31} has an approximately 100 x better angular resolution than our in-situ data.  The c-GeTe peaks were essentially resolution limited (not shown). This rules out lattice microstrain in the cubic phase, and indicates a dynamic origin for the nanoscale inhomogenieties inferred from the data presented below. The refined values of the symmetry inequivalent Ge-Te bond lengths are shown in Fig. 2a as a function of temperature. These can be seen to converge on a (symmetry imposed) average value at $T_{C}$, in a manner consistent with earlier work.\\
\begin{figure}[tb!]
\begin{center}
\includegraphics[scale=0.7]{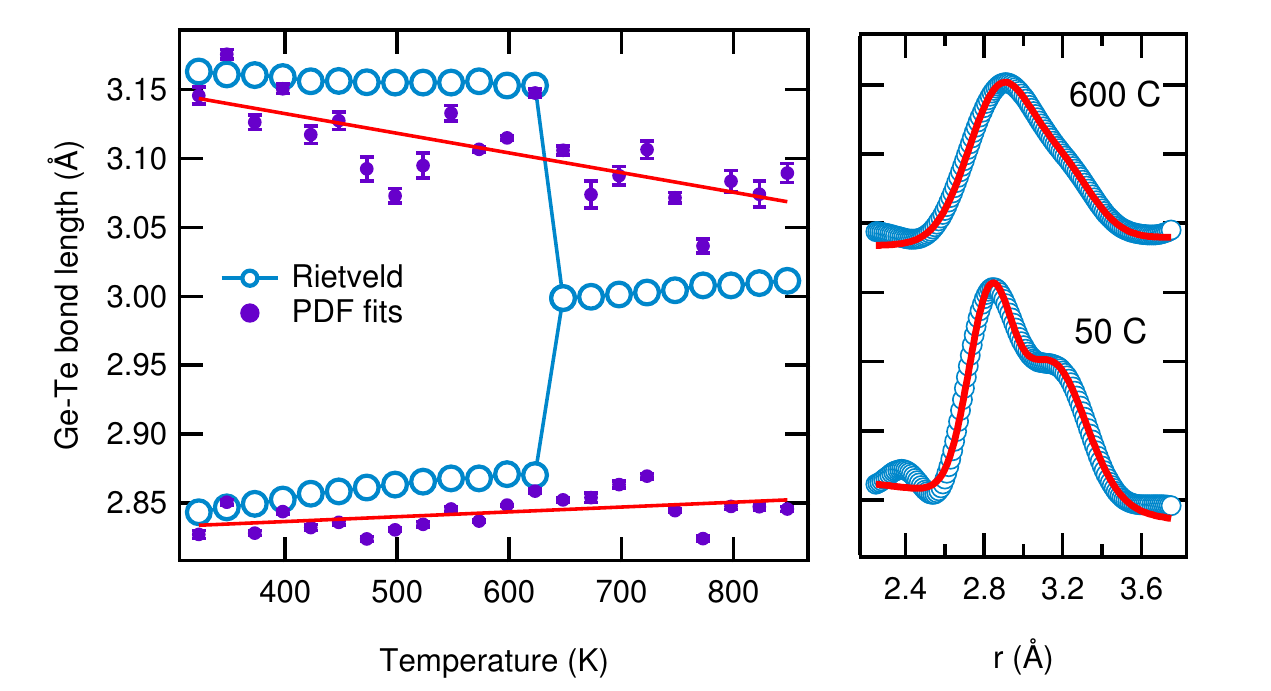}
\caption{(color online) (a) Nearest-neighbour Ge-Te bond distances extracted from Rietveld refinements (open circles) and Gaussian fits (solid circles) to the pair distribution function data; b), c), Nearest-neighbour peak splitting observed in the pair distribution functions in the $R3m$~and $Fm\bar{3}m$~phases. The lines are the results of fitting two Gaussians and a sloping background.}
\label{Fig1}
\end{center}
\end{figure}
The PDFs which we have calculated from our high energy x-ray scattering data explicitly contain the diffuse scattering signal, not just the information encoded in the Bragg reflections. Upon Fourier transforming our data, it is immediately apparent that a long and a short Ge-Te distance persist to the highest measured temperatures (Fig. 2). A simple model-independent fitting of the first coordination sphere bond lengths from the pair distribution function data highlights the dichotomy between local and average structure measurements. As shown in Fig. 2b and 2c, two contributions to the Ge-Te distances can be distinguished at all temperatures. Gaussian fits as a function of temperature show that these only slowly converge, with an estimated transition temperature well above the melting point ($\sim$1700 K). The resultant energy scale (0.15 eV), is consistent with strong covalent bonding \cite{Li2RuO3}, and explains why the SOJT distortion persists even into the molten state \cite{melt}. Clearly this degree of freedom is of secondary relevance to the mechanism of the ferroelectric transition, which has a characteristic energy scale of around 55 meV.\\
The above results contain little information beyond that contained in previously published works. In order to go further, we must establish the sensitivity of our $local$~structure measurements to the $average$~structure as measured by e.g. diffraction. It is well established that structure refinement against the pair distribution function should tend to the average structure at larger values of $r$~ in real space \cite{Li2RuO3,LaMnO3}. This is due to the Fourier relationship between reciprocal and real space. Crudely put, we can reduce the sensitivity to the broad diffuse scattering by choosing an appropriate frequency window, delimited by low-$r$ and high-$r$ cutoffs. We therefore performed real space refinements of the GeTe crystal structure in the range 30 $<$ $r$~$<$50 \AA~ as a function of temperature. As described elsewhere \cite{res}, we used a damping parameter calculated using a CeO$_{2}$~standard to account for the intrinsic instrumental resolution. This yielded the unit cell volumes shown using solid symbols in Fig. 1b. This nicely reproduces the Rietveld result, proving that the data content of our PDFs extrapolates between the local and average structure limits.\\
\begin{figure}[tb!]
\begin{center}
\includegraphics[scale=0.57]{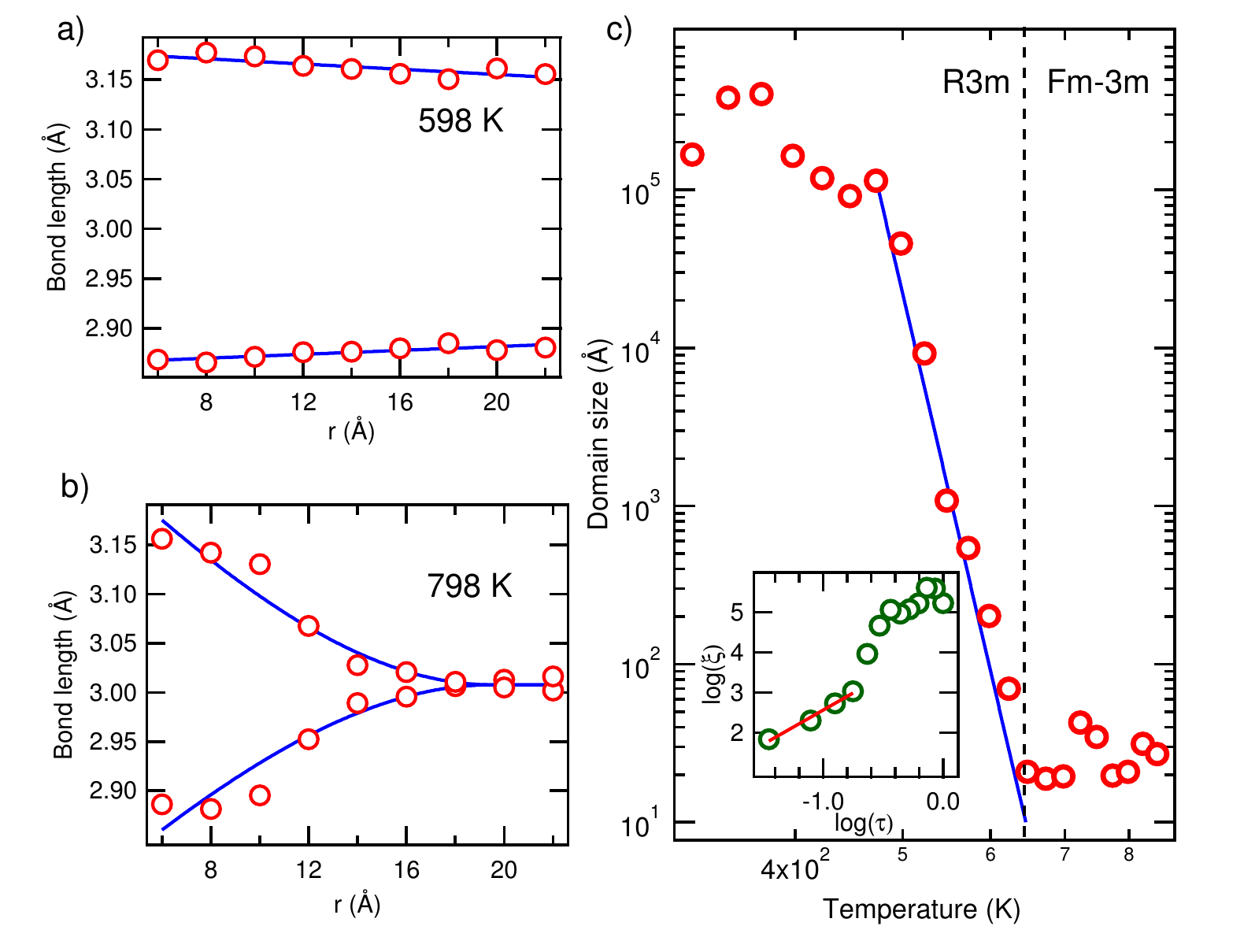}
\caption{(color online) (a) Real-space dependence of the nearest neighbour bond distance as fitted to the pair distribution function at 598 K. Lines show fits of a spherical particle envelope used to extract the correlation length; b) As in panel a), except data shown are for the $Fm\bar{3}m$~phase at 798 K; c) Extracted correlation length as a function of temperature. A power law fit is shown in the $R3m$~region. The inset shows the power law scaling as a function of reduced temperature, $\tau=\frac{(T_{c}-T)}{T_{c}}$, where $T_{c}$ is 650 K.}
\label{Fig1}
\end{center}
\end{figure}
\begin{figure}[tb!]
\begin{center}
\includegraphics[scale=0.7]{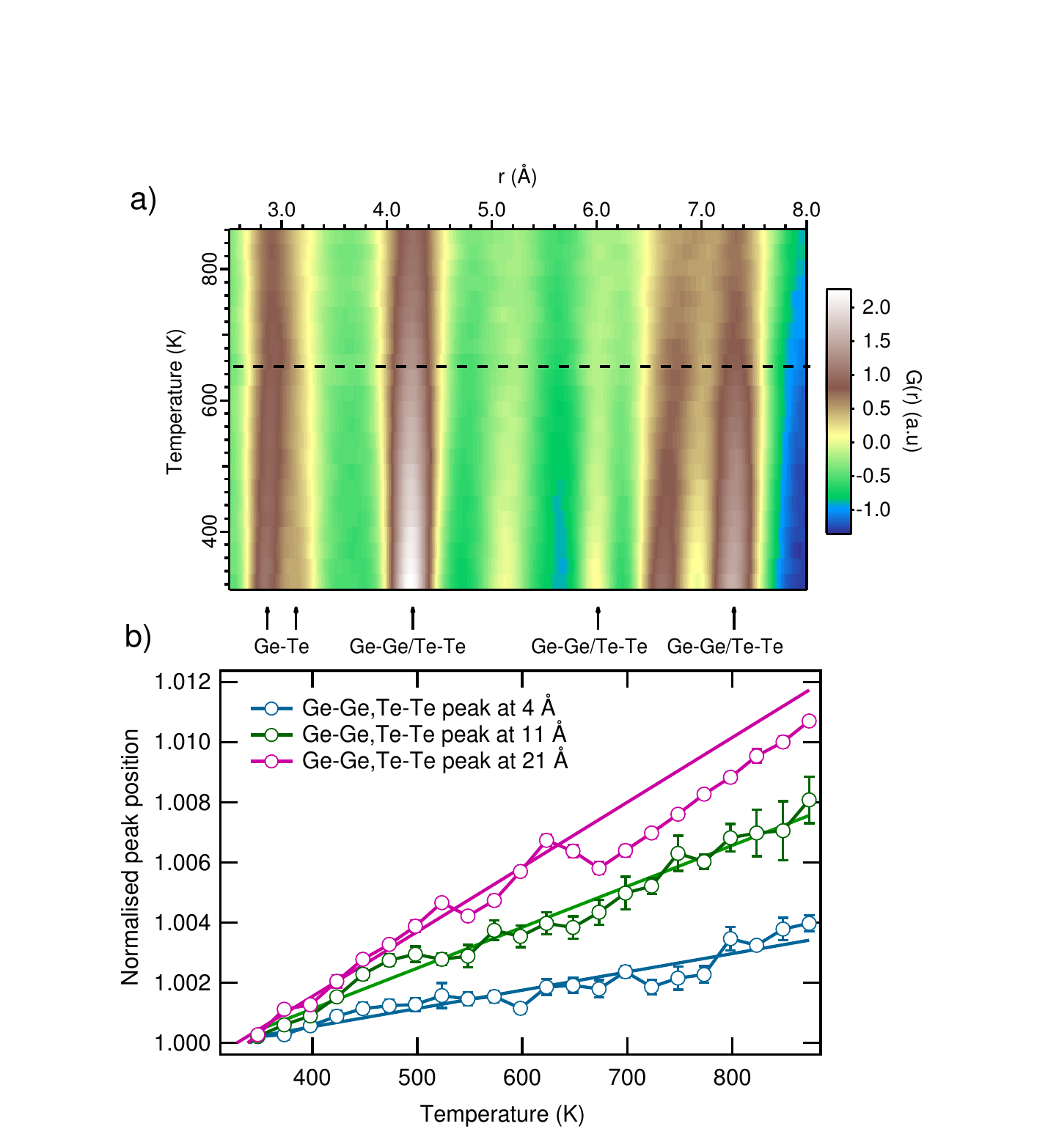}
\caption{(color online) a) Temperature dependence of the low-$r$~pair distribution function of GeTe, several important distances are highlighted. The dashed line shows the temperature at which the bulk $R3m$~to $Fm\bar{3}m$~transition occurs; b) Temperature dependence of peak positions in the pair distribution function. Note the crossover to behaviour mirroring that of the unit cell volume at higher values of $r$. Lines are guides to the eye, and error bars are shown, but are often smaller than the symbol sizes.}
\label{Fig1}
\end{center}
\end{figure}
To recap, the PDF of GeTe: i) Shows essentially no change in nearest-neighbour distances at the structural phase transition and ii) Tends to the known average structure result at longer distances. The key question is therefore what happens on $intermediate$~length scales upon warming through the ferroelectric phase transition. To answer this question, we performed $r$-dependent refinements of the $R3m$ crystal structure for all temperatures. This model was found to describe all data sets satisfactorily at low-$r$. By plotting the Ge-Te bond-lengths as a function of $r$-max and fitting the spherical domain envelope function~\cite{LaMnO3}, we extracted a correlation length for the local order. Note that the fitting was done in a self-consistent manner, using a single variable to describe the correlation length for both bond lengths. Close to room temperature, the correlation length is essentially infinite, yet at higher temperatures, it begins to decrease. As shown in Fig. 3c, the correlation length, $\xi$, decays as $\xi\propto T^{-\upsilon'}$. This is reminiscent of the soft-modes observed experimentally with neutron~\cite{Tapan2} and Raman~\cite{Raman} spectroscopies, which also show power law behaviour near the transition. Our observations are strikingly different however, as the correlation length does not tend to zero, but reaches a constant level of~$\xi\sim$20~\AA~in the cubic phase. This is essentially unchanged until the highest temperature we have measured (900 K), and we speculate that these nanoscale domains persists all the way up to the melting point of GeTe at 998 K. The melt is known to contain~\cite{melt} the same local SOJT distortion as found at all temperatures here, and thus may well correspond to the loss of coherence in the $\sim$20 \AA~domains discovered herein. Note that a simple analysis of the power law scaling (inset Fig. 3c) provides a critical exponent of $\upsilon'\sim$1.7(1). This is far from the mean field value of $\frac{1}{2}$, and shows that the transition is far from a conventional soft-mode driven transition.\\
Finally, we wish to comment on the negative thermal expansion observed at the structural phase transition (Fig.1 b). This phenomena can be driven by many diverse mechanisms, such as gross changes~\cite{elecNTE} in electronic structure, or purely by low-energy vibrational degrees of freedom. The latter is common in framework materials with low coordination numbers~\cite{zwo1,zwo2}, yet is more difficult to justify in a close-packed material like GeTe, which also shows abrupt, rather than continuous NTE. While our results do not provide a definite mechanism for the volume collapse, we can associate it with the 2 nm distorted domains discovered above. A colour map of the low-$r$~region of the pair distribution functions is shown in Fig. 4a as a function of temperature. No major anomalies can be observed, which might indicate a reduction in atomic distances. In fact, model independent fitting (Fig. 4b) shows that e.g. the Ge-Ge/Te-Te peak at 4.1\AA~ smoothly moves to high-$r$. We performed this analysis at several positions, attempting to find peaks with low multiplicities. This becomes challenging as the length scale increases, as the number of distances probed in the PDF is $\propto 4\pi r^{2}$. Nevertheless, clear signatures of the volume collapse are apparent as the length scale increases to $\sim$~20 \AA, matching the size of the distorted domains discovered above. We note that tellurium-based alloys generally show NTE even in the liquid state. This has been convincingly attributed to the interplay between entropy and volume~\cite{newPRL}. Although our measurements are a time average of the thermally occupied states in the system, and we are $a~priori$~unable to distinguish between static and dynamic disorder, we speculate that the latter is equally important in 'crystalline' GeTe.\\
In summary, the above results show that the structural phase transition found at 650 K in the prototypical phase change material GeTe is of an exotic nature. The energy scale relevant to the nearest-neighbour SOJT distortion is apparently relevant only at temperatures far exceeding the melting point. In contrast, a coarse-grained order-disorder model, made up of ~ 2 nm domains, can explain the results of both local and average structure probes. Future measurements sensitive to the length scales probed here, as well as dynamics, will be useful to further understand this important parent material.\\

We thank Gavin Vaughan for useful discussions, and Adrian Hill for assistance with the measurement on ID31.  The ESRF is acknowledged for access to instrumentation. SJLB was supported by US DOE, Office of Science, Office of Basic Energy Sciences (DOE-BES), under Contract No. DE-SC00112704.

\end{document}